\documentclass{emulateapj}
\usepackage[figuresright]{rotating}
\accepted{for publication by AJ}

\shorttitle{Young clusters in M31} \shortauthors{Ma et al.}

\begin{document}
\title{AGE AND MASS STUDIES FOR YOUNG STAR CLUSTERS IN M31 FROM SEDS-FIT}

\author{
Song Wang\altaffilmark{1,2,3}, Jun Ma\altaffilmark{1,3}, Zhou Fan\altaffilmark{1,3}, Zhenyu Wu\altaffilmark{1,3}, Tianmeng Zhang\altaffilmark{1,3}, Hu Zou\altaffilmark{1,3}, Xu Zhou\altaffilmark{1,3}}

\altaffiltext{1}{National Astronomical Observatories, Chinese Academy of Sciences, Beijing, 100012, P. R. China;
majun@nao.cas.cn}

\altaffiltext{2}{Graduate University of Chinese Academy of Sciences, A19 Yuquan Road, Shijingshan
District, Beijing 100049, China}

\altaffiltext{3}{Key Laboratory of Optical Astronomy, National Astronomical Observatories, Chinese
Academy of Sciences, Beijing, 100012, China}

\begin{abstract}
In this paper, we present photometry for young star clusters in M31, which are selected from Caldwell et al. These star clusters have been observed as part of the Beijing--Arizona--Taiwan--Connecticut (BATC) Multicolor Sky Survey from 1995 February to 2008 March. The BATC images including these star clusters are taken with 15 intermediate-band filters covering 3000--10000 \AA.
Combined with photometry in the {\sl GALEX} far- and near-ultraviolet, broad-band $UBVRI$, SDSS
$ugriz$, and infrared $JHK_{\rm s}$ of Two Micron All Sky Survey, we obtain their accurate spectral energy distributions (SEDs) from $1538-20000$ \AA. We derive these star clusters' ages and masses by comparing their SEDs with stellar population synthesis models. Our results are in good agreement with previous determinations. The mean value of age and mass of young clusters ($<2$ Gyr) is about 385 Myr and $2\times 10^4~{M_\odot}$, respectively. There are two distinct peaks in the age distribution, a highest peak at age $\sim$ 60 Myr and a secondary peak around 250 Myr, while the mass distribution shows a single peak around $10^4~{M_\odot}$. A few young star clusters have two-body relaxation times greater than their ages, indicating that those clusters have not been well dynamically relaxed and therefore have not established the thermal equilibrium. There are several regions showing aggregations of young star clusters around the 10 kpc ring and the outer ring, indicating that the distribution of the young star clusters is well correlated with M31's star-forming regions. The young massive star clusters (age $\leq 100$ Myr and mass $\geq 10^4~{M_\odot}$) show apparent concentration around the ring splitting region, suggesting a recent passage of a satellite galaxy (M32) through M31 disk.
\end{abstract}


\section{INTRODUCTION}
Star clusters are considered as important tracers for understanding the formation and evolution of their host galaxies \citep{san10}. Star cluster systems have been traditionally separated into two populations--globular clusters and open clusters (GCs and OCs)--on their ages, masses, metallicities, and positions. However, more recent studies have discovered that the distinction between GCs and OCs becomes increasingly blurred \citep[see][for details]{perina10}.

\citet{gk52} listed photometric colors and magnitudes for star clusters in Magellanic Clouds (MCs) and the Fornax dwarf system and divided them into two groups. They found that star clusters in blue group have central condensation properties similar to those of the red group, which were considered as GCs, however they could not be identified with the Galactic OCs. \citet{hodge61} termed 23 clusters in the Large Magellanic Cloud (LMC)--differing from GCs in their relative youth and OCs in their richness and shape--as ``young populous clusters'', which were called ``young massive clusters'' (YMCs) or ``blue luminous compact clusters'' (BLCCs) by \citet{Fusi05}. Actually, the blue integrated colors for a cluster may be influenced by several factors , such as poor metallicity (the luminosity of the horizontal branch), young age (the position of the main-sequence turnoff stars), and some exotic stellar populations (e.g., blue stragglers, Wolf-Rayet stars). However, several studies \citep[e.g.,][]{Williams01,Beasley04} have reached similar conclusions that the exceedingly blue colors of BLCCs are a direct consequence of their young ages \citep[see][for details]{Fusi05}.

M31 is the largest galaxy in the Local Group, and has a large number of star clusters, including young clusters having been studied by many authors. \citet{bohlin88,bohlin93} listed 11 objects in M31 classified as blue clusters using the UV colors,
most of which have been proved to be young clusters \citep{Fusi05,cald09,perina09,perina10}, except for
B133 and B145, which were stated as a star and an old GC \citep{cald09}, respectively. \citet{cald09, cald11} derived ages and masses for a large sample of young clusters, and found that these star clusters are less than 2 Gyr old, and most of them have ages between $10^8$ and $10^9$ yr and masses ranging from $2.5 \times 10^2~{M_\odot}$ to $1.5 \times 10^5~{M_\odot}$. These authors also stated that the young star clusters in M31 show a range of structures, most of which have low concentrations typical of OCs in the Milky Way (MW), however, there are a few with high concentrations similar to the MW GCs. \citet{vanse09} carried out a survey of compact star clusters in the southwest part of M31, and suggested a rich intermediate-mass star cluster population in M31, with a typical age range of 30 Myr $-$ 3 Gyr, peaking at $\sim$ 70 Myr. In order to ascertain the properties of the BLCCs, \citet{perina09,perina10} performed an image survey for 20 BLCCs lying in the disk of M31 using the Wide Field and Planetary Camera-2 (WFPC2) on the {\it Hubble Space Telescope} ({\it HST}). In addition, another key aim of this {\it HST} survey was to determine the fraction of contamination of BLCCs by asterisms, since \citet{Cohen05} suggested that a large fraction of the putative BLCCs may in fact be just asterisms. \citet{Cohen05} presented the resulting $K'$ images of six very young or young star clusters in M31 observed with the Keck laser guide star adaptive optics system, and indicated that the four youngest out of these six objects are asterisms. However, \citet{cald09} presented a conclusion that these four objects are true clusters based on spectra. The {\it HST} images \citep{perina09,perina10} showed that nineteen of the twenty surveyed candidates are real star clusters, and one (NB67) is a bright star. \citet{barmby09} measured surface brightness profiles for 23 bright young star clusters using images from the WFPC2, including the sample clusters of \citet{perina09,perina10}, and derived the structural properties by fitting the surface brightness profiles to several structural models. The authors stated that the sample young clusters are expected to dissolve within a few Gyr and will not survive to become old GCs, and that young star clusters in M31 and MCs follow the same fundamental plane relations as old GCs of M31, MCs, the MW and NGC 5128, regardless of their host galaxy environments. \citet{johnson12} presented a M31 stellar cluster catalog utilizing the Panchromatic Hubble Andromeda Treasury survey data, which will cover $\sim1/3$ of M31 disk with multiple filters and allow the identification of thousands of star clusters.

The large population of young star clusters reflect a high efficiency of cluster formation, possibly triggered by a current interaction event between M31 and its satellite galaxy \citep{Gordon06, block06}, suggesting that young star clusters should be associated with the star-forming (SF) regions of M31. \citet{fan10} found that young clusters ($<$ 2 Gyr) are spatially coincident with M31's disk, including the 10 kpc ring and the outer ring \citep{Gordon06}. Although these authors also found the young star clusters in the halo of M31, all of the clusters outside of the optical disk of M31 are old, globular clusters \citep[see][for details]{perina11}. \citet{kang12} stated that most of young star clusters' kinematics have the thin, rotating disk component \citep[see also][]{rey07}. The young star clusters' distribution has a distinct peak around $10-12$ kpc from the center in M31 disk, and some young star clusters show concentration around the 10 kpc ring splitting regions near M32 and most of them have systematically younger ages ($< 100$ Myr). \citet{kang12} also stated that the young star clusters show a spatial distribution similar to OB stars, UV SF regions, and dust, all of which are important tracers of disk structures.

Several criteria were developed for selecting young clusters from the integrated spectrum and colors. \citet{Fusi05} comprehensively studied the properties of 67 very blue and likely YMCs in M31 selected according to their color $[(B-V)_0\leq 0.45]$ and/or the strength of $H\beta$ spectral index ($\rm H\beta \geq 3.5$ \AA). \citet{Peacock10} presented a catalog of M31 GCs based on images from the SDSS and the Wide Field CAMera on the United Kingdom Infrared Telescope and selected a population of young clusters with a definition of $[(g-r)_0<0.3]$. \citet{kang12} published a catalog of M31 young clusters ($\leq 1$ Gyr) and supported the selection criteria $[(NUV-r)_0\leq 2.5]$ and $[(FUV-r)_0\leq 3.0]$ \citep{bohlin93,rey07}. These criterions may play important roles in distinguishing young from old clusters for those whose ages cannot be derived accurately.

The formation and disruption of young star clusters represent a latter-day example of the hierarchical formation of galaxies \citep{fall04}. Motivated by that, we decided to describe some basic properties of young star clusters in M31, such as positions, distributions of ages and masses, correlations of the ages and masses with structure parameters, which may provide important information about the processes involved in their formation and disruption.

In this paper, we will provide photometry of a set of young star clusters in M31 using images obtained with the Beijing--Arizona--Taiwan--Connecticut (BATC) Multicolor Sky Survey Telescope. By comparing the observed  SEDs with the {\sc galev} simple stellar population (SSP) models, we derive their ages and masses. This paper is organized as follows. In Section 2 we present the BATC observations of the sample clusters, the relevant data-processing steps, and the {\sl GALEX} (FUV and NUV), optical broad-band, SDSS $ugriz$ and 2MASS NIR data that are subsequently used in our analysis. In Section 3 we derive ages and masses of the sample clusters. A discussion on the sample young clusters ($<$ 2 Gyr) will be given in Section 4.
Finally, we will summarize our results in Section 5.

\section{SAMPLE OF STAR CLUSTERS, OBSERVATIONS, AND DATA REDUCTION}

\subsection{Sample of Star Clusters}

The sample of star clusters in this paper is selected from \citet{cald09,cald11}, who presented a series of studies of M31 young and old clusters, respectively. We selected 178 young clusters given in \citet{cald09,cald11}, and fortunately, all the young clusters have been observed with 15 intermediate-band filters of the BATC photometric system. However, there are 42 clusters for which we cannot obtain accurate photometric measurements with different reasons as following, a) some clusters have one or more nearby very bright objects; b) some clusters are very close to other objects; c) some clusters are very faint and the signal-to-noise ratio (SNR) is low; d) some clusters are superimposed onto a bright background; e) some clusters are superimposed onto a strongly variable background.
In addition, There are several remarkable clusters with ``adhered'' \citep{vanse09} objects in our images, such as M088 and its neighbor M089, and G099 and C037-G099x \citep[see also][]{narbutis08}.
In a previous paper, \citet{ma11} presented the SEDs in 15 intermediate-band filters of the BATC photometric system for one YMC VDB0-B195D and determined its age and mass by comparing its SEDs with the theoretical evolutionary population synthesis models. Thus, here we will analyze the multicolor photometric properties of the remaining 135 clusters.

\subsection{Archival Images of the BATC Sky Survey for M31 Field}

The M31 field is part of a galaxy calibration program of BATC Multicolor Sky Survey. The BATC program uses the 60/90 cm Schmidt Telescope at the Xinglong Station of the National Astronomical Observatories, Chinese Academy of Sciences (NAOC). This system includes 15 intermediate-band filters,
covering a range of wavelength from 3000 to 10000 \AA~\citep[see][for details]{fan96}. Before February 2006, a Ford Aerospace $2{\rm k}\times2{\rm k}$ thick CCD camera was applied, which has a pixel size of 15 $\mu\rm{m}$ and a field of view of $58^{\prime} \times 58^{\prime}$, resulting in a resolution of $1''.67~\rm{pixel}^{-1}$. After February 2006, a new $4{\rm k}\times4{\rm k}$ CCD with a pixel size of 12 $\mu$m was used, with a resolution of $1''.36$ pixel$^{-1}$ \citep{fan09}. We obtained 143.9 hours of imaging of the M31 field covering about 6 square degrees, consisting of 447 images, through the set of 15 filters in five observing runs from 1995 to 2008, spanning 13 years \citep[see][for details]{fan09,wang10}.

Figure 1 shows the spatial distribution of the sample clusters and the M31 fields observed with the BATC multicolor system, in which a box only indicates a field view of $58^{\prime}$ $\times $ $58^{\prime}$ of the thick CCD camera. All the sample star clusters are indicated with black dots, with high-accuracy coordinates from \citet{cald09,cald11}, which are based on the images from the Local Group Galaxies Survey \citep{massey06} and the Digitized Sky Survey.

\begin{figure}
\figurenum{1} \resizebox{\hsize}{!}{\rotatebox{0}{\includegraphics{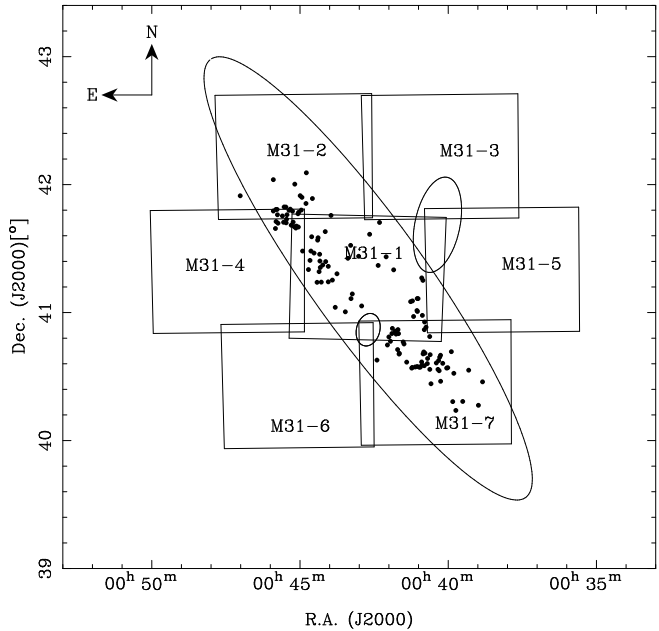}}}
\caption{Spatial distribution of 135 sample star clusters indicated by black dots. A box represents a field view of $58^{\prime}$ $\times $ $58^{\prime}$. The large ellipse
is the M31 disk/halo boundary as defined by \citet{rac91}; the two small ellipses are the $D_{25}$ isophotes of NGC 205 (northwest) and M32 (southeast).} \label{fig:fig1}
\end{figure}

\subsection{Integrated Photometry of the Sample Star Clusters}

We processed all the CCD images to apply standard procedures including bias subtraction and flat-fielding using an automatic data reduction software named PIPELINE I, developed for the BATC Multicolor Sky Survey of the CCD images. BATC magnitudes are defined and obtained in a similar way as for the spectrophotometric AB magnitude system \citep{ma09b}. In order to improve the image quality, multiple images of the same filter were combined to one, on which the magnitudes of the sample star clusters were determined. The absolute flux of the combined images in the central field of M31 (M31-1 in Figure 1) was calibrated using observations of standard stars, while the absolute flux of the combined images of the M31-2 to M31-7 fields was calibrated based on secondary standard transformations using the M31-1 field \citep[see][for details]{fan09}.

We performed standard aperture photometry of our sample objects using the {\sc PHOT} routine in {\sc DAOPHOT} \citep{stet87}. To ensure that we adopted the most appropriate photometric radius that included all light from the objects, we used 9 different aperture sizes (with radii of $r_{\rm ap}=2.0, 2.5, 3.0, 3.5, 4.0, 4.5, 5.0, 5.5, 6.0$ pixels for the old CCD, while for the new CCD, the radii were given with pixels corresponding to the same arcsecs on the old CCD) to determine the magnitude. We also checked the aperture radii carefully on the images by visual examination in order not to include the light from extraneous objects. The local sky background was measured in an annulus with an inner radius of $r_{\rm ap}+1.0$ pixel and a width of 4.0 pixels for the old CCD, and with an inner radius of $r_{\rm ap}+1.0$ pixel and a width of 5.0 pixels for the new CCD, respectively.

There are 40 clusters in this paper which are in common with our series of previous papers
\citep{jiang03, ma06, ma09b, fan09, wang10}, and the photometric data of these 40 clusters were also
derived by these studies. We found that most photometric data obtained here are in good agreement with those obtained by previous studies. We checked the images of those clusters with photometric discrepancy, and found most of them were loaded near the bulge (B091) or in some disk regions with bright or variable background
(e.g., B210, M020, M023). The different choice of the aperture for photometry and the annulus for
background caused the discrepancy.

The SEDs for the sample clusters in M31 are listed in Table 1. Columns (1) gives the cluster names. Columns (2) to (16) present the magnitudes in the 15 BATC passbands. The $1\sigma$ magnitude uncertainties from {\sc DAOPHOT} are listed for each object on the second line
for the corresponding passbands. For some objects, the magnitudes in some filters could not be obtained because of low SNR. We should remind that magnitudes with an uncertainty larger than 0.3 will not be used in
the following analysis, although they are listed in Table 1. Columns (17) is the photometric aperture adopted in this paper.

\subsection{{\sl GALEX} UV, Optical Broad-band, SDSS, and 2MASS NIR Photometry}

As our series of papers has pointed out, accurate and numerous photometric points can derive accurate ages of star clusters \citep{degrijs03, anders04}. \citet{kavirag07} stated that the UV photometry is powerful for age estimation of young stellar populations, and the combination of UV photometry with optical observations enables one to break the age-metallicity degeneracy.
\citet{jong96} and \citet{anders04} showed that the age-metallicity degeneracy can be partially broken by adding NIR photometry to optical colors \citep[see][and references therein]{ma09b}. Several previous studies \citep{bh00,gall04,rey07,Peacock10,kang12} have provided magnitudes for star clusters in different passbands, which will be used to estimate ages of sample star clusters in this paper.

The latest Revised Bologna Catalogue of M31 GCs and candidates (hereafter RBC v.4) \citep{gall04,gall06,gall07,gall09} includes {\sl GALEX} (FUV and NUV) fluxes from \citet{rey07}, optical broad-band, 2MASS NIR magnitudes for 2045 objects.
For $UBVRI$ magnitudes given in RBC v.4, the relevant photometric uncertainties are not listed. Therefore, we adopted the original $UBVRI$ measurements of \citet{bh00} as our preferred reference, including their published photometric errors. For the remaining objects,
the $UBVRI$ magnitudes from RBC v.4 were adopted, with the photometric uncertainties set following \citet{gall04}, i.e., $\pm 0.08$ mag in $U$ and $\pm 0.05$ mag in $BVRI$
\citep{ma09b}.

In RBC v.4, the 2MASS $JHK_{\rm s}$ magnitudes were transformed to CIT photometric system \citep{gall04}. However, we needed the original 2MASS $JHK_{\rm s}$ data to compare the observed SEDs with the SSP models, so we reversed the transformation using the equations given by \citet{Carpenter01}. There were no magnitude errors for $JHK_{s}$ bands in RBC v.4, and we obtained them by comparing the photometric data with Figure 2 of \citet{Carpenter01}, in which the photometric error was shown as a function of magnitude for stars brighter than their observational completeness limits \citep{ma09b, wang10}. In addition, since RBC v.4 provided $JHK_{\rm s}$ magnitudes only for a small number of sample star clusters, we adopted $JHK_{\rm s}$ magnitudes from the 2MASS-6X-PSC catalog, with 6 times the normal exposure of 7.2 s on most fields of M31 \citep{nantais06}. There are 3 kinds of magnitudes given by 2MASS-6X-PSC catalog, the ``default'' magnitude, the $r=4''$ aperture and $r=10''$ aperture magnitudes. We found that the $r=4''$ aperture magnitudes agree well with the magnitudes in RBC v.4, while the other 2 kinds of magnitudes have large discrepancy with magnitudes in RBC v.4. However, for the $r=4''$ aperture magnitudes, when $JHK_{\rm s}$ magnitudes are fainter than $m = 16$ mag, the dispersions are considerable. In this paper, we preferentially adopted 2MASS $JHK_{\rm s}$ magnitudes in RBC v.4. For the remaining star clusters, we adopted the $r=4''$ aperture magnitudes in the 2MASS-6X-PSC catalog when the magnitudes brighter than $m=16$ mag.

\citet{Peacock10} performed SDSS $ugriz$ photometry for 1595 M31 clusters and cluster candidates
using the program SExtractor on drift scan images
of M31 obtained by the SDSS 2.5-m telescope, which
are on the AB photometric system \citep[see][for details]{bertin96}. We found that there was a magnitude offset ($\geq 0.5 \rm~mag$) between SDSS and $UBVRI$ for 35 objects by transforming between the $ugriz$ and $UBVRI$ bands following the transformations from \citet{Jester05}. Because the SDSS $ugriz$ magnitudes provided a more homogeneous set of photometric measurements,
we adopted the SDSS magnitudes and abandoned $UBVRI$ magnitudes for these objects in the following analysis. These 35 objects are flagged with small ``a'' in Column 1 of Table 2.

\citet{kang12} presented a catalog of 700 confirmed star clusters in M31, providing the most extensive and updated UV integrated photometry on the AB photometric system based on {\sl GALEX} imaging, superseding the UV photometry published by \citet{rey07}, which were included by RBC v.4. Therefore, we used the magnitudes of the FUV and NUV from \citet{kang12} as the UV photometry in our following SED fitting process.

We listed the {\sl GALEX}, optical broad-band, SDSS $ugriz$, and 2MASS NIR photometry of the sample clusters in Table 2 (Columns 2 to 16), where the photometric errors are listed for each object on the second line for the corresponding passbands. As we discussed above, the magnitudes with an uncertainty larger than 0.3 will not be used in the following analysis.

\subsection{Comparison with Previously Published Photometry}

To check our photometry, we transformed the BATC intermediate-band system to the broad-band
system using the relationships between these two systems derived by \citet{zhou03}:

\begin{equation}
B=m_{d}+0.2201(m_{c}-m_{e})+0.1278\pm0.076 \quad \mbox{and}
\end{equation}

\begin{equation}
V=m_{g}+0.3292(m_{f}-m_{h})+0.0476\pm0.027.
\end{equation}

$B$-band photometry can be derived from the BATC $c, d$, and $e$ bands, while $V$-band magnitude can be obtained from the BATC $f, g$, and $h$ bands. Figure 2 shows a comparison of the $B$ and $V$ photometry of our M31 sample objects with previous measurements from \citet{bh00} (circles) and \citet{gall04} (triangles).

There are several objects with larger offsets ($\Delta m > 0.5$), shown with black solid marks in Figure 2 (M045 in the top panel; B200D, SK036A, and SK068A in the bottom panel). The SNRs of M045, B200D, and SK036A are low, and SK068A is superimposed onto a bright background, thus we cannot derive accurate photometries for these four star clusters.

The mean $B$ and $V$ magnitude differences--in the sense of this paper minus others--are $\langle \Delta B \rangle =0.002 \pm 0.213$ mag and $\langle \Delta V \rangle =0.081 \pm 0.209$, i.e., there is no system offset between our magnitudes and previous determinations.

\begin{figure}
\figurenum{2}\resizebox{\hsize}{!}{\rotatebox{-90}{\includegraphics{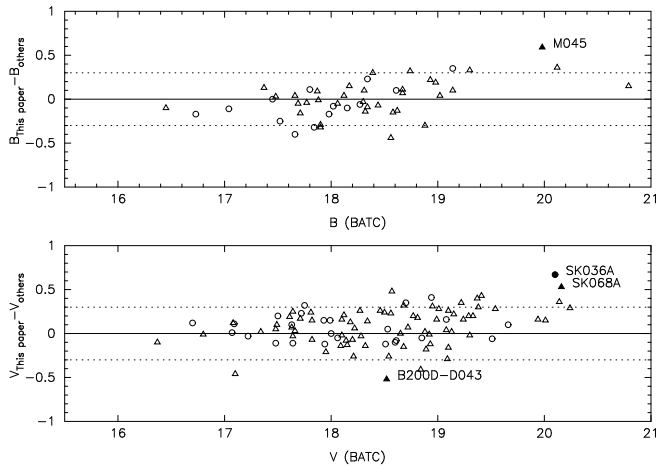}}}
\caption{Comparison of our newly obtained star cluster photometry with previous measurements by
\citet{bh00} (circles) and \citet{gall04} (triangles). The dashed lines enclose $\pm 0.3$ mag in $B$ and $V$. The black filled circles and triangles indicate the objects with photometry offset $> 0.5$ mag with \citet{bh00} and \citet{gall04}, respectively.}
\label{fig:fig2}
\end{figure}

\subsection{Reddening Values}

We required independently determined reddening values to estimate ages of the sample clusters robustly and accurately. Here we used \citet{kang12} and \citet{cald09} as our reference. \citet{kang12} derived reddening values from three ways: 1) mean reddening values from available literature \citep{bh00, fan08, cald09, cald11}; 2) median reddening values of star clusters located within an annulus at each 2 kpc radius from the center of M31, for these star clusters there were no available reddening values in the literature; 3) for star clusters at distances larger than 22 kpc from the center of M31, the foreground reddening value of $E(B-V)=0.13$ was adopted. Because all of our sample clusters have a projected galactocentric radius smaller than 22 kpc, the reddening values for them were derived from the first two methods.
The reddening value of cluster LGS04131.1\_404612 was not given by \citet{kang12}, so we adopted the value $E(B-V)=0.20$ from \citet{cald09}. Its reddening uncertainty was simply adopted half of the reddening value, i.e., $\sigma_{E(B-V)}=0.10$. We noticed that, for star cluster B449, the reddening value determined by \citet{kang12} is very different with the value determined by \citet{cald11} ($\Delta E(B-V) = 1.14$). We treated both age and reddening value as free parameters, and determined the reddening value to be $E(B-V)=0.10$ which was in good agreement with \citet{cald11}. So, in this paper ,we adopted $E(B-V)=0.13$ in \citet{cald11}. The reddening uncertainty for B449 was adopted to be 0.07. Column 4 of Table 4 lists the reddening values adopted for the sample clusters, while Column 5 lists the methods for deriving the reddening values ($\rm flag=1$ and $2$ indicate that the reddening values were obtained by the first and second method in Kang et al. 2012, respectively; $\rm flag=3$ indicates that the reddening values were from Caldwell et al. 2009, 2011, only for LGS04131.1\_404612 and B449).

\section{AGE AND MASS DETERMINATION}

\subsection{Stellar Populations and Synthetic Photometry}

In order to determine the ages and masses of the sample star clusters, we compared their SEDs with theoretical stellar population synthesis (SPS) models. The SSP models of {\sc galev} \citep[e.g.,][]{kurth99,schulz02,anders03} were adopted \citep{ma09b, wang10} in this paper, which are based on the Padova isochrones (with the most recent versions using the updated Bertelli 1994 isochrones, including the thermally-pulsing asymptotic giant-branch [TP-AGB] phase), and a \citet{salp55} stellar initial mass function (IMF) with a lower-mass limit of $0.10~{M_\odot}$ and the upper-mass limit between 50 and 70 $M_\odot$ depending on metallicity. The full set of models span the wavelength range from 91 {\AA} to 160 $\mu$m. These models cover ages from $4 \times 10^6$ to $1.6 \times 10^{10}$ yr, with an age resolution of 4 Myr for ages up to 2.35 Gyr, and 20 Myr for greater ages. The {\sc galev} SSP models include five initial metallicities, $Z=0.0004, 0.004, 0.008, 0.02$ (solar metallicity), and 0.05.

Since our observational data were integrated luminosities through our set of filters, we convolved the {\sc galev} SSP SEDs with the {\sl GALEX} FUV and NUV, broad-band $UBVRI$, SDSS $ugriz$, BATC, and 2MASS $JHK_{\rm s}$ filter response curves to obtain synthetic ultraviolet, optical, and NIR photometry for comparison. The synthetic $i{\rm th}$ filter magnitude in the AB magnitude system can be computed as

\begin{equation}
m_i=-2.5\log\frac{\int_{\nu}F_{\nu}\varphi_{i} (\nu){\rm d}\nu}{\int_{\nu}\varphi_{i}(\nu){\rm
d}\nu}-48.60,
\end{equation}
where $F_{\nu}$ is the theoretical SED and $\varphi_{i}$ is the response curve of the $i{\rm th}$ filter of the corresponding photometric systems. Here, $F_{\nu}$ varies with age and metallicity.

\subsection{Fits}

We used a $\chi^2$ minimization test to determine which {\sc galev} SSP models are most compatible with the observed SEDs, following

\begin{equation}
\chi^2=\sum_{i=1}^{n}{\frac{[m_{\nu_i}^{\rm intr}-m_{\nu_i}^{\rm mod}(t)]^2}{\sigma_{i}^{2}}},
\end{equation}
where $m_{\nu_i}^{\rm mod}(t)$ is the integrated magnitude in the $i{\rm th}$ filter of a theoretical SSP at metallicity $Z$ and age $t$, $n$ is the number of the filters used for fitting, $m_{\nu_i}^{\rm intr}$ represents the intrinsic integrated magnitude in the same filter and

\begin{equation}
\sigma_i^{2}=\sigma_{{\rm obs},i}^{2}+\sigma_{{\rm mod},i}^{2}+(R_{\lambda_i}*\sigma_{\rm red})^2
+\sigma_{{\rm md},i}^{2}.
\end{equation}
Here, $\sigma_{{\rm obs},i}$ is the observational uncertainty, and $\sigma_{{\rm mod},i}$ is the uncertainty associated with the model itself, for the $i{\rm th}$ filter. \citet{charlot96} estimated the uncertainty associated with the term $\sigma_{{\rm mod},i}$ by comparing the colors obtained from different stellar evolutionary tracks and spectral libraries. Following
\citet{ma07,ma09a,ma09b,ma11,ma12} and \citet{wang10}, we adopted $\sigma_{{\rm mod},i}=0.05$ mag in this paper.
$\sigma_{\rm red}$ is the uncertainty in the reddening value, and $R_{\lambda_i}= A_{\lambda_i}/E(B-V)$,
where $A_{\lambda_i}$ is taken from \citet{car89}, $R_V= A_V/E(B-V)=3.1$, and $\sigma_{{\rm md},i}$ is the uncertainty of the distance modulus, which is always 0.07 from $(m-M)_0=24.47\pm0.07$ mag
\citep{McConnachie05}.

\citet{perina09,perina10} determined ages for 20 possible YMCs in M31 with metallicity as a free parameter of their fit. Their results showed that most YMCs in M31 were best fitted with solar metallicity model. \citet{cald09} claimed that it seemed likely the young star clusters have supersolar abundances. In this paper, the {\sc galev} models of solar metallicity ($Z=0.02$) were used to fit the intrinsic SEDs for all the sample clusters here. As an example, we presented the fitting for some sample clusters in Figure 3. During the fitting, we found that, for a small number of clusters, some photometric data cannot be fitted with any SSP models. We therefore did not use these deviating photometric data points to obtain the best fits. These deviating photometric data points are the $a$ magnitude of M070, $b$ magnitude of M040, $k$ magnitude of V133, $m$ magnitude of M082 and M101, $n$ and $p$ magnitudes of B195, $J$ magnitude of B118D, and $H$ and $K_{\rm s}$ magnitudes of M091. We also noticed that, for some star clusters (B091, B305, B319, B392, B458, B480, B484, and DAO69), the photometries in $JHK_{\rm s}$ bands show obvious offsets from {\sl GALEX} FUV and NUV bands \citep[see also][]{kang12}. Considering that the UV photometry is powerful for age estimation of young stellar populations \citep{kavirag07}, we would adopt the {\sl GALEX} FUV and NUV photometries in the SED fitting, and abandon the $JHK_{\rm s}$ magnitudes for these clusters.

\begin{figure}
\figurenum{3} \resizebox{\hsize}{!}{\rotatebox{-0}{\includegraphics{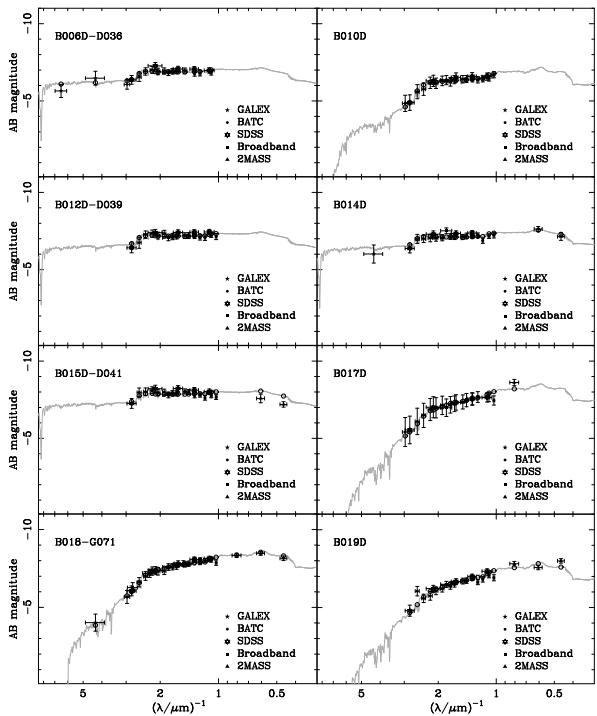}}}
\vspace{-2.cm}
\caption{Best-fitting
integrated SEDs of the {\sc galev} SSP models shown in relation to the intrinsic SEDs for our sample star clusters. The photometric data points are represented by the symbols with error bars (vertical error bars for uncertainties and horizontal ones for the approximate wavelength coverage for each filter). Open circles represent the calculated magnitudes of the model SEDs for each filter.}
\label{fig:fig3}
\end{figure}

The masses of the sample star clusters were determined sequentially. The {\sc galev} models provide absolute magnitudes in 77 filters for SSPs of $10^6~{M_\odot}$,
including 66 filters of the {\it HST}, Johnson $UBVRI$ \citep{landolt83}, Cousins $RI$ \citep{landolt83}, and $JHK$ \citep{bb88} systems. The difference between the intrinsic and model absolute magnitudes provides a direct measurement of the cluster mass, in units of $10^6~{M_\odot}$ \citep[see][for details]{ma11}. We transformed the 2MASS $JHK_{\rm s}$ magnitudes to the photometric system of \citet{bb88} using the equations given by \citet{Carpenter01}, and estimated masses of the clusters using magnitudes in all of the $UBVRI$ and $JHK_{\rm s}$ bands. The masses of clusters obtained based on the magnitudes in different filters were different,
therefore, we averaged them as the final cluster mass.

The masses of 22 clusters were not derived, because the $UBVRIJHK_{\rm s}$ magnitudes cannot be used: (1) there are no $UBVRIJHK_{\rm s}$ magnitudes (LGS04131.1\_404612); (2) there are no $JHK_{\rm s}$ magnitudes and the $UBVRI$ magnitudes were abandoned because of the discrepancy with the SDSS $ugriz$ magnitudes (e.g., B195); (3) the $UBVRI$ magnitudes were abandoned because of the discrepancy with the SDSS $ugriz$ magnitudes and the $JHK_{\rm s}$ magnitudes were abandoned because of the discrepancy with the {\sl GALEX} UV magnitudes (e.g., B319). The ages and masses of the sample clusters obtained in this paper are listed in Table 4.

\subsection{Comparison with Previous Determinations}

In this paper, we determined ages and masses for 135 star clusters by comparing their multicolor photometries with theoretical SPS models. These star clusters were from \citet{cald09,cald11}, who presented a series of studies of M31 young and old clusters, respectively.

As discussed in Section 2.3, there are 40 clusters in this paper which are in common with our series of previous papers, and the ages of 27 clusters were also derived by our series of studies \citep{jiang03,fan06,ma06,ma09b,wang10}.
In the study of \citet{jiang03}, the SSP models of Bruzual \& Charlot (G. Bruzual \& Charlot 1996, unpublished) were used; in the studies of \citet{fan06} and \citet{ma06}, the SSP models of \citet{bru03} were used. In addition, in these three studies, we used only the BATC photometries in 13 passbands, and did not include UV data, which is powerful tool for age estimation of young stellar populations \citep[see][and reference therein]{kang12}. So, we re-estimated the ages for these young star clusters with more photometric date including UV data, and with the same SSP models as \citet{ma09b} and \citet{wang10} used. In the studies of \citet{ma09b} and \citet{wang10}, we used the metallicities obtained by \citet{bh00} and \citet{per02} when estimating the ages of star clusters. The metallicities in \citet{bh00} and \citet{per02} were determined from the Lick indices which were calibrated from the Galactic old GCs. However, \citet{Fusi05} claimed that young star clusters are probably not so metal-poor as deduced from the metallicities obtained by \citet{per02}, and concluded that $G-$band line strength tends to underestimate [Fe/H] values in \citet{per02} by more than 1 dex \citep[see also][]{kang12}. As a result, most of ages obtained in \citet{ma09b} and \citet{wang10} are older than those obtained in this paper. So, we re-estimated the ages for the young star clusters in \citet{ma09b} and \citet{wang10} with the solar metallicity SSP models.

In addition, nine clusters (B476, BH11, M026, M040, M045, M053, M057, M058, and M070) were estimated to be older than 2 Gyr in this paper, which are considered to be old star clusters \citep[e.g.,][]{cald09}. Since this paper focused on young clusters, we would not consider these nine old clusters in the following analysis. However, we pointed out that the ages of six of these nine star clusters were estimated to be younger than 2 Gyr by previous studies: the age of B476 was estimated to be 1.2 Gyr by \citet{cald09} (however, 7.1 Gyr by Fan 2010); the age of BH11 was estimated to be 1.6 Gyr by \citet{vanse09}; the age of M040 was estimated to be $\sim130$ Myr by \citet{fan10} and $\sim320$ Myr by \citet{kang12}; the age of M053 was estimated to be 1 Gyr by \citet{cald09} and $\sim140$ Myr by \citet{fan10}; the age of M058 was estimated to be $\sim160$ Myr by \citet{fan10}; the age of M070 was estimated to be 1.2 Gyr by \citet{cald09} and $\sim810$ Myr by \citet{fan10}. The ages of these nine clusters derived by previous papers and here are listed in Table 3. There are three clusters with ages of 4 Myr, which is the lowest age limit of {\sc galev} models. Clusters KHM31-37 and V133 were estimated slightly older by other authors \citep{cald09,kang12}, while B196D was estimated slightly younger ($\sim2$ Myr) by \citet{cald09}.

In Figure 4, the estimated ages for the young ($<2$ Gyr) star clusters in this paper were compared with those from previous studies
\citep[e.g.,][]{Beasley04,vanse09,fan10,perina10,cald09,cald11,kang12}.
The star clusters with ages of 4 Myr obtained in this paper are drawn with open squares in Figure 4. There are five clusters in common between \citet{Beasley04} and this paper. We can see that the ages of \citet{Beasley04} are in good agreement with ours.
\citet{vanse09} estimated ages of star clusters located in the southern disk of M31 with $UBVRI$ SED-fitting. There is an obvious offset between their estimated ages and ours, which is caused by some large scatters.
If the three clusters with ages greater than 3 Gyr given by \citet{vanse09}, which are drawn with arrows
in Figure 4, are not included, the systematic offset can be reduced to be $-0.25$ Gyr.
\citet{fan10} estimated ages of star clusters in M31 with multi-band ($UBVRIJHK_{\rm s}$) SED-fitting,
and their results agree well with ours, with a small offset ($\sim0.02$ Gyr). \citet{perina09,perina10}
determined ages of 20 possible YMCs in M31 by comparing the observed color magnitude diagrams and the isochrones of different metallicities and ages of \citet{Girardi02}, and estimated masses of these clusters based on the Maraston's SSP models of solar metallicity and \citet{salp55} and \citet{Kroupa01} IMFs and the IR magnitudes in the 2MASS-6X-PSC catalog. In general, the ages obtained in this paper are in good agreement with the determinations by \citet{perina10}, with a small deviation ($\sim0.06$ Gyr). There is a small systematic offset ($\sim-0.10$ Gyr) between \citet{cald09,cald11} and this paper, i.e. the ages obtained by \citet{cald09,cald11} are larger than the ages obtained here. \citet{perina10} also found the systematic offset as the ages obtained by \citet{cald09} are larger than the ages obtained by \citet{perina10}. \citet{perina10} suggested that this offset is caused by the super-solar metallicity models ($Z = 0.04$) adopted by \citet{cald09} when they determined the ages of star clusters.
\citet{kang12} derived ages for young clusters by fitting the multi-band photometry with model
SEDs, and their results are in good agreement with ours.

\begin{figure}
\figurenum{4}
\resizebox{\hsize}{!}{\rotatebox{-90}{\includegraphics{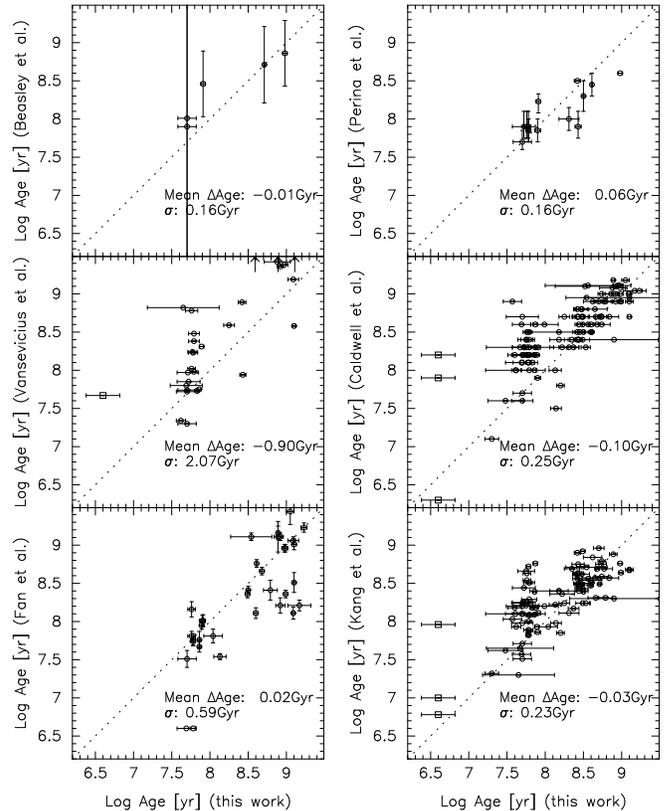}}}
\caption{Comparison of the ages obtained here with those obtained by previous works:
\citet{Beasley04}, \citet{vanse09}, \citet{fan10}, \citet{perina10}, \citet{cald09,cald11}, and \citet{kang12}. In each panel, the mean value of the age differences (ours minus other study) is given, with its standard deviation ($\sigma$). The error bars of ages from each study are also shown.}
\label{fig:fig4}
\end{figure}

In Figure 5, we compared the masses of clusters obtained in this paper with those from previous studies
\citep[e.g.,][]{Beasley04,vanse09,fan10,perina10,cald09, cald11,kang12}.
The masses of two (KHM31-37 and V133) of the three clusters with ages of 4 Myr, which are drawn with open squares in Figure 5, were derived to be lower than $10^3~{M_\odot}$. The masses estimated in this paper are in good agreement with those estimated by \citet{Beasley04}, \citet{perina10}, \citet{cald09,cald11}, and \citet{kang12}. There is an obvious offset between \citet{vanse09} and this paper, which is mainly caused by some scatters. The cluster
with the largest discrepancy is B335, with a mass estimate of $\sim5\times 10^5~{M_\odot}$ by \citet{vanse09}, and $\sim1.3\times 10^5~{M_\odot}$ in this paper. When B335 is excluded, the offset can be reduced to be ($\sim-0.6\times 10^4~{M_\odot}$). The masses estimated by \citet{fan10} are slightly less than those estimated here, with an offset of $\sim1.5\times 10^4~{M_\odot}$.

\begin{figure}
\figurenum{5}
\resizebox{\hsize}{!}{\rotatebox{-90}{\includegraphics{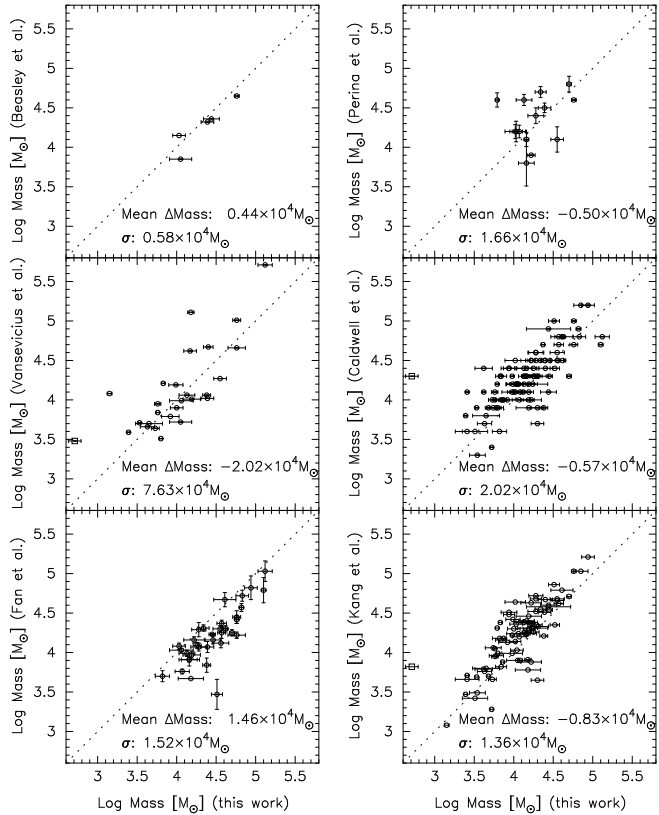}}}
\caption{Comparison of the masses obtained here with those obtained by previous works:
\citet{Beasley04}, \citet{vanse09}, \citet{fan10}, \citet{perina10}, \citet{cald09,cald11}, and \citet{kang12}. In each panel, the mean value of the mass differences (ours minus other study) is given, with its standard deviation ($\sigma$). The error bars of masses from each study are also shown.}
\label{fig:fig5}
\end{figure}

\section{DISCUSSION OF YOUNG STAR CLUSTERS}

\subsection{Position}

Figure 6 shows the number, ages and masses of young star clusters ($<2$ Gyr) as a function of projected radius from
the center of M31, adopted at $\rm \alpha_0=00^h42^m44^s.30$ and $\rm \delta_0=+41^o16'09''.0$ (J2000.0) following \citet{hbk91} and \citet{per02}. In the top panel, the histogram for the radial distribution of young star clusters shows clearly two peaks at $4-7$ kpc and $9-11$ kpc, while in the middle panel and bottom panel, wide age and mass distributions can be seen in these two peak regions.

\begin{figure}
\figurenum{6}
\resizebox{\hsize}{!}{\rotatebox{-90}{\includegraphics{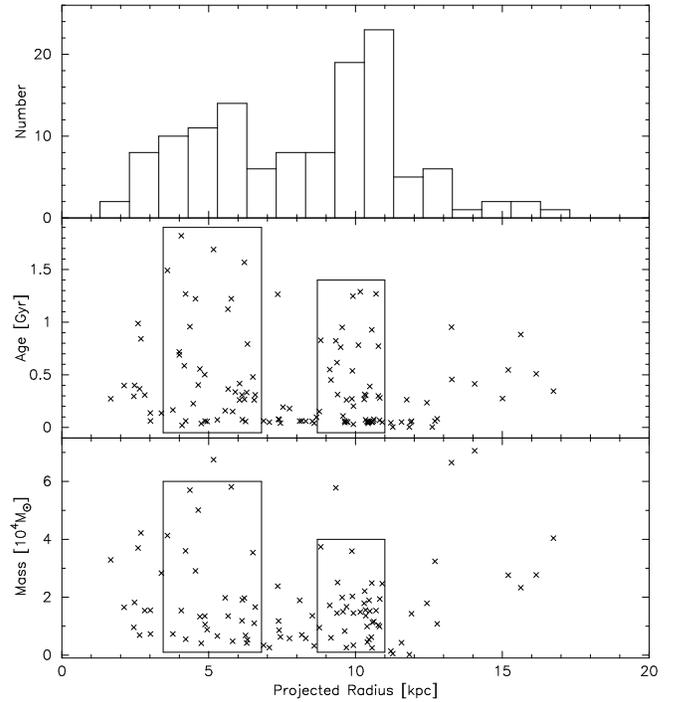}}}
\caption{
($Top~panel$) Number histogram of young star clusters against projected radius. ($Middle~panel$) Age versus projected radius for sample young clusters. ($Bottom~panel$) Mass versus projected radius for sample young star clusters. The open rectangles show the two peaks around $4-7$ kpc and $9-11$ kpc of the radial distribution for young star clusters.}
\label{fig:fig6}
\end{figure}

\citet{kang12} presented the radial distribution of clusters against the distance from the center of M31,
and found that the young clusters show two peaks around $10-12$ kpc and $13-14$ kpc. They also found that the UV SF regions show two distinct peaks: a main peak at $\sim16$ kpc and a secondary peak around 11 kpc. In addition, a small peak at $5-8$ kpc in the distribution of ages of UV SF regions against the projected radius \citep[see Figure 19 of][]{kang12} can be clearly found. We argued that the peak at $4-7$ kpc obtained in this paper should be associated with the peak at $5-8$ kpc for the UV SF regions, while the peak at $9-11$ kpc obtained in this paper correlate with the well-known 10 kpc ring \citep{Gordon06}.

Figure 7 displays the spatial distribution and radial distribution of the M31 young clusters with
different age bins: (a) $t<0.1$ Gyr; (b) 0.1 Gyr $\leq t<$ 0.4 Gyr; (c) 0.4 Gyr $\leq t<$ 1 Gyr; (d) 1 Gyr $\leq t<$ 2 Gyr. In the top panel, young star clusters in different age ranges are drawn with different marks. The inner, solid ellipse and the dashed contour represent the 10 kpc ring and the outer ring from \citet{Gordon06} based on infrared observations with the Multiband Imaging Photometer for Spitzer (MIPS) instrument on the {\it Spitzer Space Telescope}, respectively. The 10 kpc ring was drawn with a center offset from the M31 nucleus by [$5'.5$, $3'.0$] \citep{Gordon06} with a radius of 44 arcmin (10 kpc). There are several regions drawn with open rectangles which show aggregations of young star clusters to different extents.
However, we should point out that these aggregations of young star clusters may be caused by the projection effect because of the inclination of M31 disk. \citet{vanse09} noted two clumps of young clusters, both of which are located in one rectangle ($\sim-13$ kpc $< X <$ $-9$ kpc and $-3$ kpc $< Y <$ 0 kpc). The star clusters in this study are spatially coincident with the disk and the rings, indicating that the distribution of the young star clusters correlates with the galaxy's SF regions, which is consistent with previous studies \citep{fan10,kang12}. In the bottom panel, the number of young star clusters in different age bins as a function of projected radial distance from the M31 center was shown. We can see that clusters younger than $0.1$ Gyr show most obvious aggregation around the 10 kpc ring.

\begin{figure}
\figurenum{7}
\resizebox{\hsize}{!}{\rotatebox{0}{\includegraphics{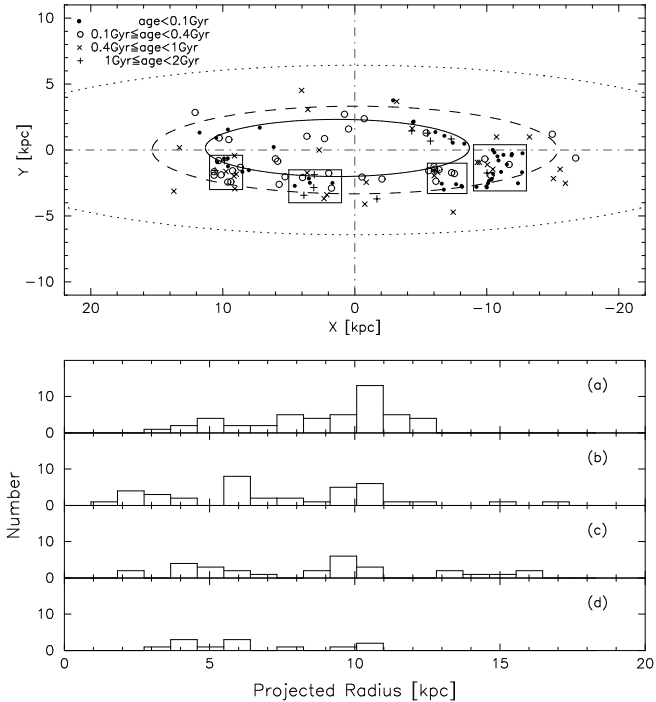}}}
\caption{Spatial distribution ($top~panel$) and radial distribution ($bottom~panel$) of M31 young star clusters with different age bins: (a) $t<$ 0.1 Gyr; (b) 0.1 Gyr $\leq t<$ 0.4 Gyr; (c) 0.4 Gyr $\leq t<$ 1 Gyr; (d) 1 Gyr $\leq t<$ 2 Gyr. The inner, solid ellipse and the dashed contour represent the 10 kpc ring and the outer ring from \citet{Gordon06}, while the dotted ellipse is the M31 disk/halo boundary as defined by \citet{rac91}. The several small rectangles show the clumps of young clusters to the extents.}
\label{fig:fig7}
\end{figure}

\citet{Gordon06} ran a number of numerical simulations of the M31--M32 and M31-NGC 205 interactions, and assumed a passage of M32 through the disk of M31 occurring 20 Myr ago, resulting in a burst of star formation that propagates outward through the disk. \citet{block06} suggested that M32 and M31 had an almost head-on collision about 210 Myr ago, and M32 passed through M31 disk again about 110 Myr ago \citep[see Figure 2 of][]{block06}, which induced two off-center rings--an inner ring with projected dimensions of $\sim$ 1.5 kpc and the 10 kpc ring. Both of the simulations recurred the 10 kpc ring and the observed split.

We divided our sample star clusters younger than 300 Myr into six groups, and showed the spatial distribution for each group in Figure 8. We can see that only star clusters with ages 50 Myr $-$ 100 Myr appear around the 10 kpc ring and the ring splitting region ($-9.5$ kpc $< X <$ $-7.5$ kpc and $-2.5$ kpc $< Y <$ $-0.5$ kpc) \citep{kang12},
indicating that 1) the 10 kpc ring may begin to form about 100 Myr ago; 2) M32 passed through the southern part of M31 disk around 100 Myr and in turn resulted in the split in the form of a hole. This appears to be consistent with the prediction by \citet{block06} of a second passage of M32 about 110 Myr ago. After the second passage, star clusters formed around the split for a long period, since there are a number of star clusters around the split with ages younger than 50 Myr. \citet{davidge12} reported that the star formation rate (SFR) of the M31 disk would be elevated greatly and quickly after an encounter event, and it would finally drop when the interstellar medium is depleted and disrupted. However, from Figure 8, we cannot find evidence of radial trend of star cluster ages \citep[see also][]{kang12,cald09}.

\begin{figure}
\figurenum{8}
\resizebox{\hsize}{!}{\rotatebox{0}{\includegraphics{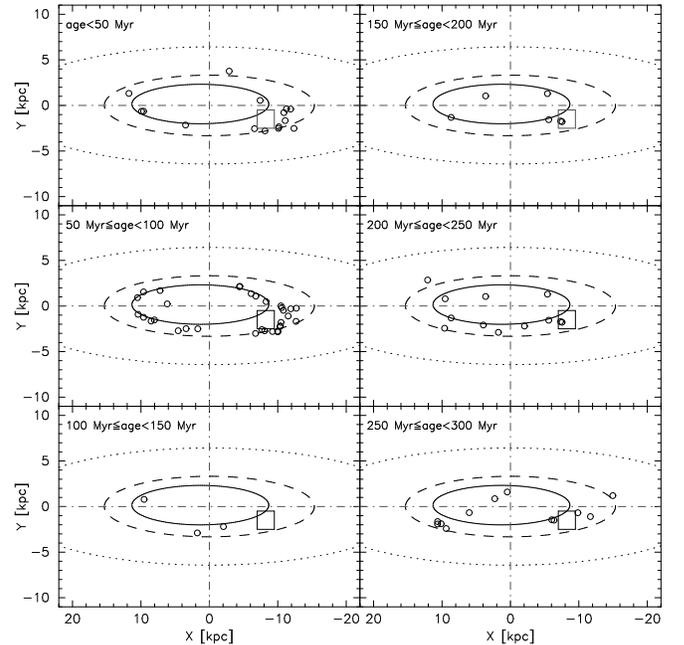}}}
\caption{Spatial distribution of six groups of M31 young star clusters younger than 300 Myr, divided with same age bin of 50 Myr. The inner, solid ellipse and the dashed contour represent the 10 kpc ring and the outer ring from \citet{Gordon06}, while the dotted ellipse is the M31 disk/halo boundary as defined by \citet{rac91}.
The small rectangle represents the ring splitting region in the southern part of M31 disk.}
\label{fig:fig8}
\end{figure}

Figure 9 shows the spatial and radial distribution of the M31 young star clusters with different mass bins:
(a) $10^2~{M_\odot} \leq M < 10^3~{M_\odot}$; (b) $10^3~{M_\odot} \leq M < 10^4~{M_\odot}$;
(c) $M \geq 10^4~{M_\odot}$. In the top panel, clusters of these three groups are drawn with different marks.
The bottom panel presents the number of young clusters in different mass bins as a function of projected radial distance from M31 center, and it shows that young clusters more massive than $10^4~{M_\odot}$ are most concentrated nearby the 10 kpc ring.

\begin{figure}
\figurenum{9}
\resizebox{\hsize}{!}{\rotatebox{0}{\includegraphics{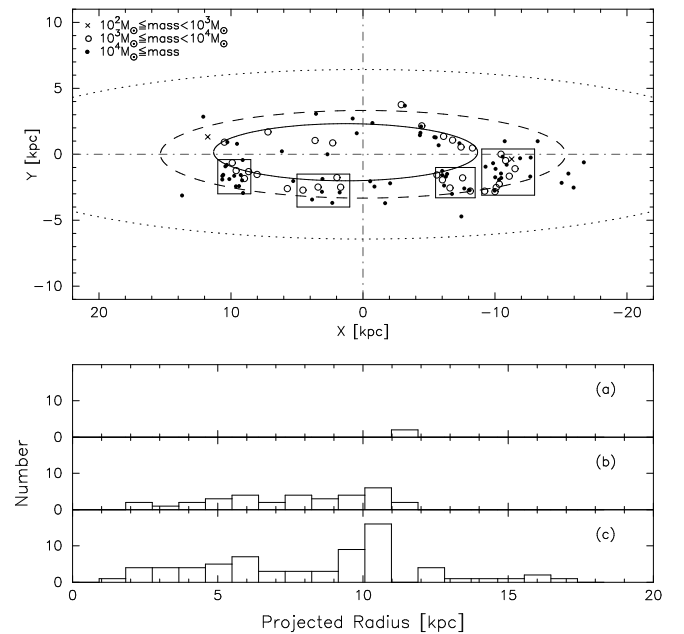}}}
\caption{Spatial distribution ($top~panel$) and radial distribution ($bottom~panel$) of M31 young star clusters with different mass bins: (a) $10^2~{M_\odot}$ $\leq M<$ $10^3~{M_\odot}$; (b) $10^3~{M_\odot}$ $\leq M<$ $10^4~{M_\odot}$; (c) $M>$ $10^4~{M_\odot}$.
The inner, solid ellipse and the dashed contour represent the 10 kpc ring and the outer ring from \citet{Gordon06}, while the dotted ellipse is the M31 disk/halo boundary as defined by \citet{rac91}. The several small rectangles show the clumps of young clusters to the extents.}
\label{fig:fig9}
\end{figure}

\subsection{Age and Mass Distribution}

Figure 10 plots the distribution of estimated ages and masses for the young star clusters. A prominent correlation can be seen that mass increases with age. There are two distinct peaks in the age histogram: a highest peak at age $\sim$ 60 Myr ($\log \rm age=7.8$) and a secondary peak around 250 Myr ($\log \rm age=8.4$). The mass distribution of the young star clusters show a single peak around $10^4~{M_\odot}$. The mean values of age and mass of young clusters are about 385 Myr and $2\times 10^4~{M_\odot}$, slightly higher than the values presented by \citet{kang12}, which are 300 Myr and $10^4~{M_\odot}$, respectively. Most of our young clusters have masses ranging from $10^{3.5}~{M_\odot}$ to $10^{5}~{M_\odot}$, which are more massive than OCs in the solar neighborhood \citep{piskunov08}, but less massive than typical GCs in the MW \citep{mm05}. The lack of young clusters more massive than $10^5~{M_\odot}$ is also noted by \citet{vanse09} and \citet{cald09}, possibly caused by a low-average SFR of M31 \citep{barmby06} or hidden by dust clouds in the disk due to the inclination angle of M31 \citep{vanse09}.

\begin{figure}
\figurenum{10} \resizebox{\hsize}{!}{\rotatebox{-90}{\includegraphics{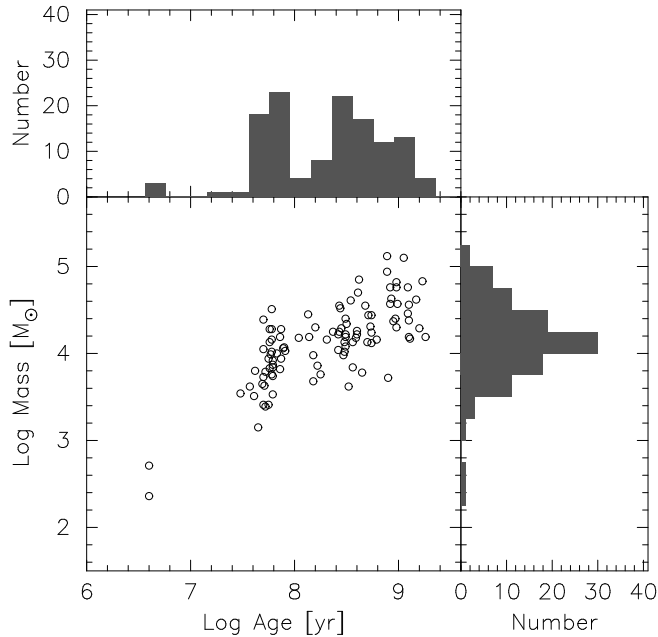}}}
\caption{Age and mass distribution of the sample young star clusters in this paper. The histograms for age and mass are presented with gray colors.}
\label{fig:fig10}
\end{figure}

\citet{portegies10} have listed three phases for the evolution of a young star cluster: 1) the first few Myr, during which the star formation activity is still proceeding and the star cluster is rich in gas; 2) a subsequent period after the first supernovae (some 3 Myr after formation), in which a young cluster is experiencing a serious loss of gas and dust, and stellar mass loss plays an important role in the cluster evolution; 3) a later stage that stellar dynamical processes dominate the cluster evolution. The dividing line between phase 2 and phase 3 may be anywhere between 100 Myr and 1 Gyr, and most of our young clusters are experiencing the phase 2 or phase 3.

\citet{cs87} presented that after 5 Gyr, both mass and galactic location are important evolutionary parameters for GCs. \citet{spitzer58} discussed the destructive effects of encounters of clusters with giant molecular clouds (GMCs), and presented that the disruption time for a star cluster varies directly with the cluster density and is about 200 Myr for a mean density of 1 $M_\odot/{\rm pc}^3$. \citet{sh58} also reported that two-body relaxation is effective at destroying low-mass clusters and this may account for the scarcity of low-mass older clusters. Actually, the two-body relaxation and the encounters with GMCs are also important processes that lead to young cluster disruption \citep[see][and references therein]{cald09}, while \citet{portegies10} presented that mass loss due to stellar evolution is the most important process in the young cluster dissolution. It is evident that star cluster mass is one key parameter in the star cluster evolution. \citet{bl03} derived an empirical relation between the disruption time and the initial mass of star clusters in the solar neighborhood, Small Magellanic Cloud (SMC), M51, and M33.
\citet{lamers05} determined a disruption time of 1.3 Gyr for a $10^4~{M_\odot}$ cluster in the solar neighborhood, while \citet{cald09} reported that most of M31 young clusters would be destroyed in the next Gyr or so, and only some massive and dense ones may survive for a longer time.

Several features are shown in Figure 10:
1) there is an obvious gap in the age distribution around 100 Myr. 2) there are few clusters older than 400 Myr (${\log \rm age=8.6}$) with mass lower than $10^4~{M_\odot}$. Although many low-mass clusters can be easy to disrupt, this gap may be caused by a selection effect. In fact, \citet{johnson12} found that the completeness of M31 ground-based sample drops precipitously at $m_{\rm F475W}>18$ ($M_{\rm F475W}>-6.5$), which is about $2\times10^4~{M_\odot}$. 3) there are few clusters more massive than $10^5~{M_\odot}$, which may be caused by a low-average SFR of M31 or the hidden by dust clouds in the M31 disk as discussed above. 4) there is a gap of clusters with very low masses ($\sim10^3~{M_\odot}$) and younger than 30 Myr (${\log \rm age=7.5}$). These clusters may be too faint to be sample objects of \citet{cald09,cald11}, indicating that our sample is not complete in these age and mass ranges \citep[see also][]{cald09}.

Figure 11 shows the age distribution in different mass intervals (top panel) and mass function in different age intervals (bottom panel). The histograms are derived using a 0.4-dex bin width with different starting values.
These distributions contain information about the formation and disruption history of star clusters \citep{fc12},
however, the interpretation of the empirical distributions of clusters depends strongly on how incompleteness affects the sample \citep{gieles07}. In the top panel, we can see an obvious gap before 40 Myr ($\log \rm age=7.6$), which is caused by a selection effect. The age distribution of the clusters does not declines monotonically, with an apparent bend around 200 Myr ($\log \rm age=8.3$). We argued that this bend near 200 Myr may be explained as a burst of cluster formation, possibly caused by a current interaction event between M31 and its satellite galaxy, such as the collision between M31 and M32 about 210 Myr ago suggested by \citet{block06}. The two decline trends starting from 40 Myr and 200 Myr reflect a rapid disruption of clusters. \citet{vanse09} noted a peak of the cluster age distribution at 70 Myr, and suggested an enhanced cluster formation episode at that epoch.
In the bottom panel, the gap in the number of clusters in the low-mass regions ($\log \rm mass < 3.5$) is apparently due to a sample incompleteness, but not physical \citep{vanse09}. The initial mass function for star clusters should be slightly steeper \citep{fc12} than what is shown here because of the short lifetimes of low-mass clusters. Recently, \citet{fc12} compared the observed age distributions and mass functions of star clusters in the MW, MCs, M83, M51, and Antennae, and found that these
distributions of clusters are similar in different galaxies. However, due to the incompleteness of our cluster sample, partly due to the exclusion of clusters that cannot derive accurate photometry, we would not give any empirical formulas of the distributions for age and mass.

\begin{figure}
\figurenum{11} \resizebox{\hsize}{!}{\rotatebox{0}{\includegraphics{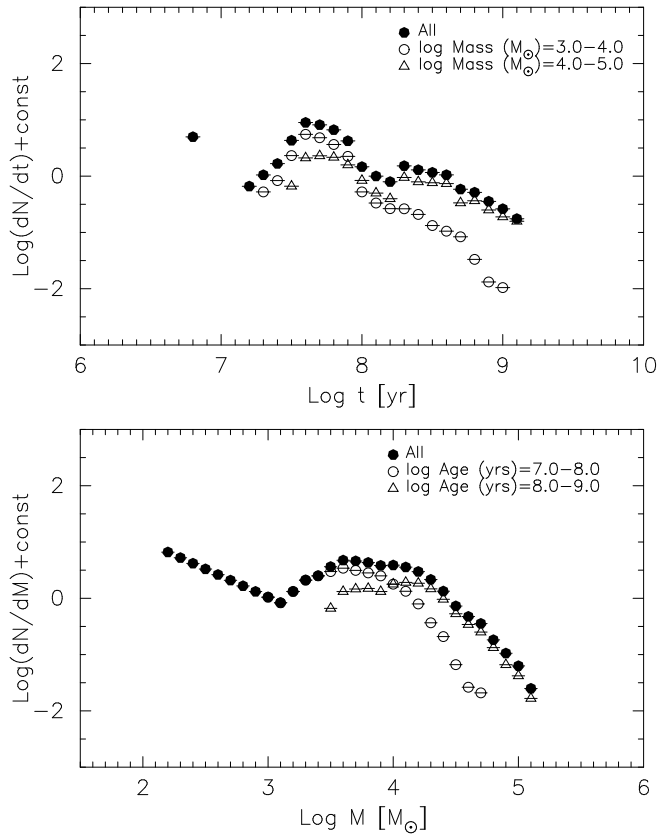}}}
\caption{Age distribution of the sample young star clusters with differen mass intervals ($top~panel$) and mass function with differen age intervals ($bottom~panel$).
The histograms are derived using a 0.4-dex bin width with different starting values.}
\label{fig:fig11}
\end{figure}

\subsection{Correlations with Structure Parameters}

In this section, we will discuss the correlations of ages and masses with structure parameters, which are derived
by King-model \citep{king66} fits for clusters in M31 \citep{bhh02, barmby07, barmby09}. Because the sample clusters are younger than 2 Gyr, the structure parameters obtained from the bluer filters are preferred \citep[see][in detail]{barmby09}. There are four clusters (B315, B319, B368, and B374) which have been studied twice by \citet{bhh02, barmby07} and \citet{barmby09}, and we would use the new results in \citet{barmby09}.

Figure 12 shows structure parameters as a function of age for young clusters in this paper. Some correlations can be seen, the concentration $c$, defined as $c\equiv\log(r_t/r_0)$, decreases with age. The trend is largely driven by clusters B342 and B368, both of which have large $c$ values (3.98 for B342 and 3.87 for B368). Both the scale radius $r_0$ and projected core radius $R_c$ increase with age. Clusters B342 and B368 have very small $r_0$ and $R_c$ values and are drawn with arrows in Figure 12 (The values of $r_0$ and $R_c$ are $\sim$ 0.014 pc for B342, while are $\sim$ 0.011 pc for B368). \citet{elson89} and \citet{elson91} discussed the trend for core radius against age, and argued that this trend may represent real evolution in the structure of clusters as they grow old, partially explained by the effect of mass segregation \citep{mg03}, or dynamical effects such as heating by black hole (BH) binaries \citep{mackey07}. \citet{wilkinson03} also demonstrated that neither large differences in primordial binary fraction nor a tidal heating due to differences in the cluster orbits could account for the observed trend. The best-fit central surface brightness $\mu_{V,0}$ shows a decreasing trend with age, and \citet{barmby09} argued that this trend may be likely due to the fading of stellar population and the increase of core radius $R_c$ with age. We also see that the central mass density $\rho_0$ decreases with age, although the scatters are great. \citet{barmby09} presented that the central mass density shows very little trend with age for both the M31 young clusters and young clusters in the MCs. There is no obvious correlation between $t_{r,h}$, the two-body relaxation time at the model-projected half-mass radius, and age. The dashed line represents the region that $t_{r,h}$ equal to age. It can be seen that most clusters (except for DAO38 and M091) have ages less than $t_{r,h}$, indicating that these young clusters have not been well dynamically relaxed. Because two-body encounters can transfer energy between individual stars and then impel the system to establish thermal equilibrium \citep{portegies10}, we argue that these young clusters have not established thermal equilibrium.

\begin{figure}
\figurenum{12} \resizebox{\hsize}{!}{\rotatebox{-90}{\includegraphics{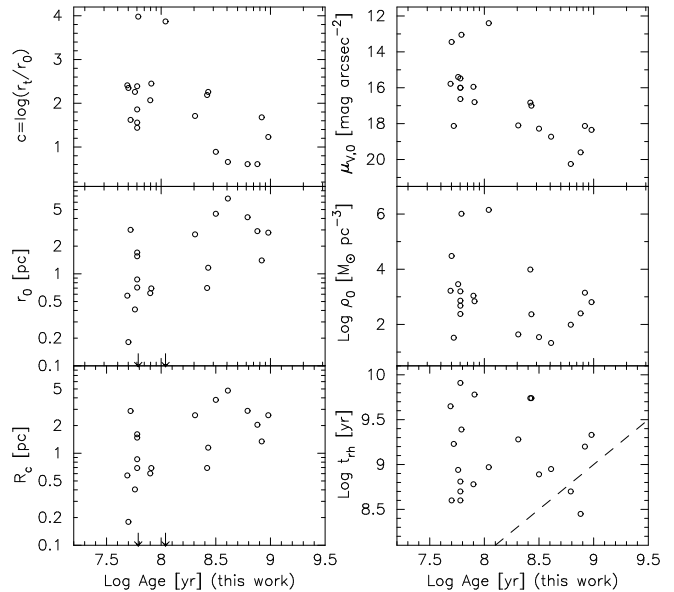}}}
\caption{Structure parameters as a function of age for the sample young star clusters in this paper.}
\label{fig:fig12}
\end{figure}

Figure 13 shows structure parameters as a function of mass for the sample young clusters. The concentration $c$ increases with mass, although the trend is much weak.
\citet{fc12} presented $c$ plotted against mass for clusters in MCs, and found that there was no correlation between $c$ and mass. Actually, we found that all the clusters in \citet{fc12} have $c$ less than 2.5, much smaller than the largest value in our sample ($\sim4$). If we do not include the two clusters B342 and B368, the correlation for $c$ with mass nearly disappear. Both $r_0$ and $R_c$ increase with mass, however, the trend is largely weaken by cluster B327, which has very small values of $r_0$ and $R_c$, but larger than those of B342 and B368 which are drawn with arrows in Figure 13. Both the central surface brightness $\mu_{V,0}$ and central mass density $\rho_0$ decrease weakly with mass, while no obvious correlation between $t_{r,h}$ and mass can be seen.

\begin{figure}
\figurenum{13} \resizebox{\hsize}{!}{\rotatebox{-90}{\includegraphics{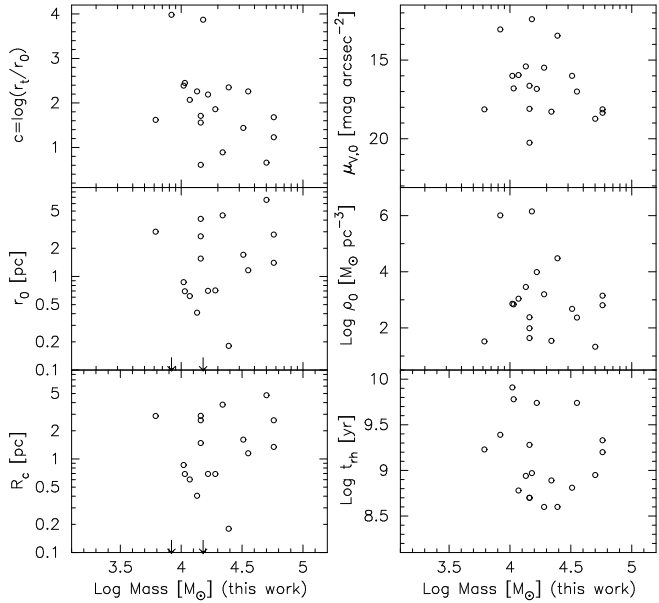}}}
\caption{Structure parameters as a function of mass for the sample young star clusters in this paper.}
\label{fig:fig13}
\end{figure}

We checked the surface brightness profiles of B327, B342, and B368 displayed in \citet{barmby09}, which have very small $r_0$ and $R_c$ and very large $\mu_{V,0}$ and  $\rho_0$, and found that these core profiles are cuspy. \citet{barmby09} concluded that the cores of these clusters did not appear to be resolved in the {\it HST}/WFPC2 images and the structural parameters for these clusters would be uncertain if the central cluster luminosity is dominated by only a few bright stars. However, if these cuspy core profiles are true integrated properties, which may be better fitted by a power-law structure model \citep[e.g.,][]{sersic68}, the three clusters may have been post core-collapse \citep[see][in detail]{tanvir12}.

\subsection{Young Massive Clusters}

YMCs are often related to the violent SF episodes triggered by galaxy collisions, mergers, and close encounters \citep{grijs07}. However, based on a sample of 21 nearby spirals, \citet{lr99} found that YMCs can exist in a wide variety of host galaxy environments, including quiescent galaxies, and that there is no correlation between the morphological type of the galaxies and their contents of YMCs. YMCs are dense aggregates of young stars, which are also expected to be the nurseries for many unusual objects, including exotic stars, binaries, and BHs \citep{portegies10}. Many studies \citep{barmby09, cald09, vanse09, Peacock10, perina10, portegies10, ma11}
that focused on M31 YMCs have derived remarkable achievements in understanding their stellar populations, structure parameters, and dynamical properties.

There are 13 YMCs in our cluster sample with a definition of age $\leq 100$ Myr and mass $\geq 10^4~{M_\odot}$ \citep{portegies10}. Figure 14 shows the spatial distribution of the 13 YMCs, while different sizes of the open circles indicate YMCs in different mass ranges.
The rectangle between the 10 kpc ring and the outer ring represents the split
in the southern part of the M31 disk, and the two black filled triangles represent M32 and NGC 205. It is not surprising to see that most of the YMCs gather around the split, indicating that there has been a high-level star formation activity, which is consistent with previous studies
\citep{Gordon06, kang12}.

\begin{figure}
\figurenum{14} \resizebox{\hsize}{!}{\rotatebox{0}{\includegraphics{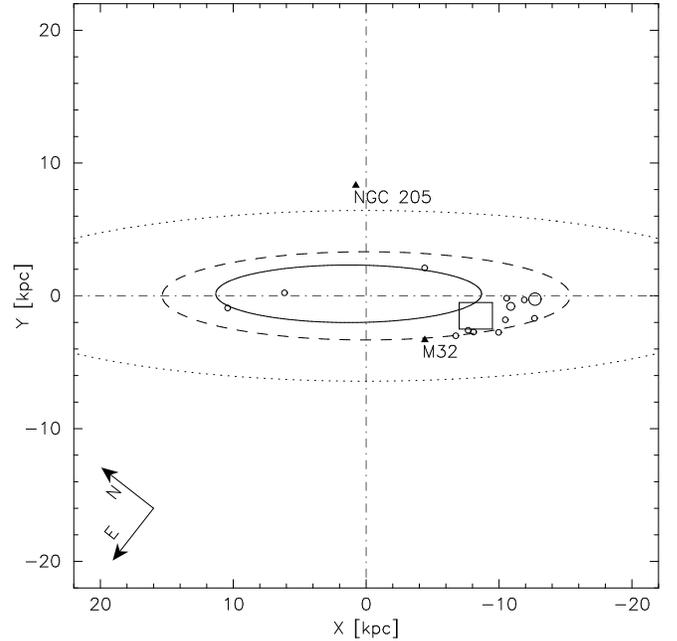}}}
\caption{Spatial distribution for YMCs drawn with different sizes of the open circles indicating different mass ranges. The inner, solid ellipse and the dashed contour represent the 10 kpc ring and the outer ring from \citet{Gordon06}, while the dotted ellipse is the M31 disk/halo boundary as defined by \citet{rac91}. The small rectangle represents the ring splitting region in the southern part of
M31 disk, and the two filled black triangles represent M32 and NGC 205.}
\label{fig:fig14}
\end{figure}

\section{SUMMARY}

In this paper, we determined the ages and masses for a sample of M31 star clusters by comparing the multicolor photometry with theoretical SPS models. Multicolor photometric data are from the {\sl GALEX} FUV and NUV, broadband $UBVRI$, SDSS $ugriz$, 15 intermediate-band filters of BATC, and 2MASS $JHK_{\rm s}$, which constitute the SEDs covering $1538-20000$ \AA.

We made a discussion on the spatial distribution, distribution of ages and masses, correlations of ages and masses with structure parameters for the sample young clusters ($<2$ Gyr). The mean value of age and mass of young clusters is about 385 Myr and $2\times 10^4~{M_\odot}$, respectively. There are two distinct peaks in the age distribution, a highest peak at age $\sim$ 60 Myr and a secondary peak around 250 Myr, while the mass distribution shows a single peak around
$10^4~{M_\odot}$. There are several regions showing aggregations of young clusters around the 10 kpc ring and
the outer ring, indicating that the distribution of the young clusters correlates well with M31's SF regions.
The ages and masses show apparent correlations with some structure parameters. We also found the correlation between core radius $R_c$ and age, which has been studied by many authors.
A few young clusters have the two-body relaxation times $t_{r,h}$ greater than their ages, indicating that they have not been well dynamically relaxed. We argued that these young clusters have not established the thermal equilibrium.

The YMCs (age $\leq 100$ Myr and mass $\geq 10^4~{M_\odot}$) show obvious aggregation around the split in the southern part of the M31 disk, suggesting a high efficiency of star formation, possibly triggered by a recent passage of a satellite galaxy (M32) through M31 disk.

\acknowledgments
We would like to thank the anonymous referee for providing rapid and thoughtful report that helped improve the original manuscript greatly. This work was supported by the Chinese National Natural Science Foundation grant Nos. 10873016, 10633020, 11073032, and 11003021, and by the National Basic Research Program of China (973 Program) No. 2007CB815403.

\newpage

%\begin{sidewaystable}
\begin{table}
\begin{center}
\small
\setlength{\tabcolsep}{0.3em} \caption{BATC intermediate-band photometry of 135
sample star clusters in M31.}
\vspace{-3mm} 
\label{t1.tab}
\begin{tabular}{lcccccccccccccccc}
\tableline\tableline
Object & $a$ & $b$ & $c$ & $d$ & $e$ & $f$ & $g$ & $h$ & $i$ & $j$ &$ k$ & $m$ & $n$ & $o$ & $p$ & $R_{\rm ap}$\\
       &(mag)&(mag)&(mag)&(mag)&(mag)&(mag)&(mag)&(mag)&(mag)&(mag)&(mag)&(mag)&(mag)&(mag)&(mag)& ($''$)  \\
\hline
 B006D-D036               &20.07   &19.46   &18.99   &18.76   &18.75   &18.61   &18.51   &18.44   &18.32   &...     &18.38   &18.29   &...     &18.10   &...     &   5.0\\
       &0.142   &0.121   &0.049   &0.056   &0.066   &0.066   &0.059   &0.063   &0.074   &...     &0.108   &0.115   &...     &0.130   &...     &    \\
 B010D                    &20.89   &20.03   &19.79   &19.28   &19.23   &18.99   &18.96   &18.78   &18.68   &18.63   &18.54   &18.57   &18.45   &18.37   &18.11   &   4.2\\
       &0.254   &0.161   &0.045   &0.038   &0.039   &0.034   &0.048   &0.040   &0.045   &0.052   &0.072   &0.059   &0.100   &0.078   &0.124   &    \\
 B012D-D039               &...     &20.14   &19.46   &19.19   &19.14   &18.83   &18.85   &18.73   &18.55   &18.47   &18.36   &18.34   &18.25   &17.98   &18.01   &   5.0\\
       &...     &0.261   &0.058   &0.087   &0.093   &0.080   &0.113   &0.114   &0.111   &0.119   &0.110   &0.117   &0.139   &0.105   &0.167   &    \\
 B014D                    &...     &19.80   &19.64   &19.40   &19.21   &19.04   &18.79   &18.73   &18.59   &18.43   &18.33   &18.16   &18.41   &18.02   &17.89   &   5.0\\
       &...     &0.158   &0.066   &0.099   &0.080   &0.074   &0.067   &0.070   &0.072   &0.074   &0.093   &0.067   &0.128   &0.097   &0.148   &    \\
 B015D-D041               &...     &19.51   &19.26   &18.92   &18.84   &18.61   &18.46   &18.34   &18.03   &17.98   &18.04   &17.86   &17.91   &17.65   &17.65   &   5.0\\
       &...     &0.135   &0.051   &0.067   &0.057   &0.053   &0.059   &0.058   &0.052   &0.061   &0.077   &0.060   &0.098   &0.068   &0.107   &    \\
 B017D                    &20.22   &19.44   &18.98   &18.51   &18.32   &18.09   &17.97   &17.87   &17.68   &17.53   &17.41   &17.22   &...     &17.14   &17.32   &   6.7\\
       &0.216   &0.154   &0.042   &0.051   &0.046   &0.042   &0.055   &0.048   &0.057   &0.059   &0.075   &0.062   &...     &0.099   &0.157   &    \\
 B018-G071                &19.81   &18.78   &18.32   &17.93   &17.88   &17.80   &17.45   &17.39   &17.20   &17.13   &16.84   &16.95   &16.79   &16.71   &16.86   &   5.8\\
       &0.274   &0.052   &0.061   &0.083   &0.048   &0.037   &0.037   &0.033   &0.026   &0.041   &0.045   &0.034   &0.048   &0.081   &0.125   &    \\
 B019D                    &...     &19.81   &20.18   &19.90   &19.48   &19.23   &18.96   &18.79   &18.51   &18.44   &18.18   &18.39   &18.15   &17.84   &17.95   &   5.0\\
       &...     &0.176   &0.139   &0.179   &0.140   &0.128   &0.124   &0.111   &0.108   &0.101   &0.099   &0.103   &0.130   &0.103   &0.163   &    \\
 B035D                    &21.08   &19.74   &19.14   &18.91   &18.67   &18.56   &18.41   &18.35   &18.13   &18.10   &18.05   &18.01   &18.18   &17.94   &18.08   &   5.0\\
       &0.399   &0.192   &0.032   &0.047   &0.040   &0.040   &0.048   &0.046   &0.053   &0.054   &0.081   &0.060   &0.117   &0.120   &0.177   &    \\
 B040-G102                &18.68   &17.73   &17.64   &17.65   &17.54   &17.48   &17.37   &17.23   &17.18   &17.18   &16.94   &17.16   &16.98   &16.93   &16.90   &   5.8\\
       &0.097   &0.020   &0.030   &0.056   &0.036   &0.027   &0.037   &0.027   &0.019   &0.038   &0.055   &0.040   &0.054   &0.112   &0.124   &    \\
\tableline
\end{tabular}
\end{center}
%\end{sidewaystable}
\end{table}

%\begin{sidewaystable}
\begin{table}
\begin{center}
\small
\setlength{\tabcolsep}{0.3em}
\caption{{\sl GALEX}, optical broad-band, SDSS,
and 2MASS NIR photometry of 135 sample star clusters in M31.}
%\vspace{3mm}
\label{t2.tab}
\begin{tabular}{lccccccccccccccc}
\tableline\tableline
Object& FUV & NUV & $U$ & $B$ & $V$ & $R$ & $I$ & $u$ & $g$ & $r$ & $i$ & $z$ & $J$ & $H$ & $K_{\rm s}$ \\
      &(mag)&(mag)&(mag)&(mag)&(mag)&(mag)&(mag)&(mag)&(mag)&(mag)&(mag)&(mag)&(mag)&(mag)&(mag)\\
\hline
 B006D-D036          &21.50   &20.87   &...     &...     &18.69   &...     &...     &19.72   &18.44   &18.30   &18.10   &17.99   &...     &...     &...     \\
       &0.100   &0.060   &...     &...     &0.050   &...     &...     &0.114   &0.069   &0.077   &0.090   &0.134   &...     &...     &...     \\
 B010D               &...     &...     &...     &...     &18.92   &...     &...     &20.80   &19.11   &18.62   &18.40   &18.27   &...     &...     &...     \\
       &...     &...     &...     &...     &0.050   &...     &...     &0.224   &0.114   &0.120   &0.127   &0.171   &...     &...     &...     \\
 B012D-D039          &...     &...     &...     &...     &19.05   &...     &...     &20.60   &19.03   &18.48   &18.13   &17.81   &...     &...     &...     \\
       &...     &...     &...     &...     &0.050   &...     &...     &0.187   &0.052   &0.068   &0.085   &0.123   &...     &...     &...     \\
 B014D               &...     &22.81   &...     &...     &18.53   &...     &...     &20.54   &19.05   &18.49   &18.14   &...     &...     &15.79   &15.72   \\
       &...     &0.380   &...     &...     &0.050   &...     &...     &0.131   &0.044   &0.057   &0.072   &...     &...     &0.133   &0.224   \\
 B015D-D041          &...     &...     &...     &...     &18.70   &...     &...     &20.39   &18.70   &18.01   &17.71   &17.42   &...     &15.92   &15.72   \\
       &...     &...     &...     &...     &0.050   &...     &...     &0.183   &0.091   &0.098   &0.108   &0.131   &...     &0.265   &0.151   \\
 B017D               &...     &...     &...     &...     &18.23   &...     &...     &20.05   &18.32   &17.77   &17.39   &17.17   &15.21   &...     &...     \\
       &...     &...     &...     &...     &0.050   &...     &...     &0.132   &0.048   &0.057   &0.066   &0.088   &0.094   &...     &...     \\
 B018-G071           &...     &22.19   &18.47   &18.25   &17.53   &17.00   &16.38   &19.39   &17.85   &17.22   &16.82   &16.63   &15.46   &14.78   &14.60   \\
       &...     &0.140   &0.080   &0.050   &0.050   &0.050   &0.050   &0.088   &0.059   &0.063   &0.067   &0.076   &0.094   &0.086   &0.112   \\
 B019D               &...     &...     &...     &...     &18.93   &...     &...     &21.14   &19.38   &18.61   &18.11   &17.59   &16.07   &15.73   &14.79   \\
       &...     &...     &...     &...     &0.050   &...     &...     &0.246   &0.043   &0.064   &0.082   &0.110   &0.112   &0.112   &0.112   \\
 B035D               &22.37   &22.95   &...     &...     &18.48   &...     &...     &20.05   &18.68   &18.27   &17.91   &17.58   &15.99   &15.35   &15.18   \\
       &0.340   &0.310   &...     &...     &0.050   &...     &...     &0.162   &0.111   &0.121   &0.135   &0.164   &0.094   &0.112   &0.130   \\
 B040-G102           &19.84   &19.30   &17.42   &17.69   &17.30   &16.92   &16.63   &18.32   &17.38   &17.19   &17.05   &16.98   &16.07   &14.96   &15.51   \\
       &0.030   &0.020   &0.046   &0.022   &0.010   &0.022   &0.022   &0.048   &0.035   &0.039   &0.045   &0.059   &0.112   &0.086   &0.130   \\
\tableline
\end{tabular}
\end{center}
%\end{sidewaystable}
\end{table}

\begin{table}
\begin{center}
\small
\setlength{\tabcolsep}{0.3em}
\caption{Age comparison for the nine clusters older than 2 Gyr obtained in this paper with previous studies.}
%\vspace{3mm}
\label{t3.tab}
\begin{tabular}{lcc}
\tableline\tableline
Object &  Log (Age)$^a$  &  Log (Age)$^b$    \\
       &    (yr)         &   (yr)            \\
\hline
B476-D074 &   $9.08^1~9.85^2$ & $  9.79\pm  0.44$ \\
BH11      &   $9.20^4$         & $  9.67\pm  0.97$ \\
M026      &   $9.93^2$         & $ 10.11\pm  0.23$ \\
M040      &   $8.11^2~8.50^3$ & $  9.38\pm  0.57$ \\
M045      &   ...              & $  9.38\pm  0.50$ \\
M053      &   $9.00^1~8.16^2$ & $  9.71\pm  0.83$ \\
M057      &   ...              & $  9.77\pm  1.03$ \\
M058      &   $8.21^2$         & $ 10.09\pm  0.29$ \\
M070      &   $9.08^1~8.91^2$ & $  9.97\pm  0.65$ \\
\tableline
\end{tabular}
\end{center}
{$^a$The age estimates are from \citet{cald09,cald11} (ref=1), \citet{fan10} (ref=2), \citet{kang12} (ref=3), and \citet{vanse09} (ref=4).}\\
{$^b$The ages obtained here.}
\end{table}

\begin{table}
\vspace{-13cm}
\small
\setlength{\tabcolsep}{0.3em}
\begin{center}
\caption{Ages, masses, and reddening values of 135 sample star clusters in M31.}
%\vspace{3mm}
\label{t4.tab}
\begin{tabular}{lcccc}
\tableline\tableline
Object &    Log (Age)  &    Log (Mass)  &  $E(B-V)$ & Flag$^a$\\
       &    (yr)      &    ($M_\odot$)  &           &       \\
\hline
 B006D-D036          & $7.78  \pm0.06  $ & $3.76  \pm0.02  $ &$0.33  \pm0.05  $ & 1     \\
 B010D               & $8.70  \pm0.07  $ & $4.13  \pm0.02  $ &$0.25  \pm0.09  $ & 1     \\
 B012D-D039          & $7.76  \pm0.08  $ & $3.83  \pm0.02  $ &$0.52  \pm0.05  $ & 1     \\
 B014D               & $7.75  \pm0.11  $ & $3.94  \pm0.07  $ &$0.50  \pm0.05  $ & 1     \\
 B015D-D041          & $7.78  \pm0.03  $ & $4.02  \pm0.13  $ &$0.65  \pm0.05  $ & 1     \\
 B017D               & $9.09  \pm0.07  $ & $4.76  \pm0.11  $ &$0.23  \pm0.18  $ & 2     \\
 B018-G071           & $8.92  \pm0.04  $ & $4.76  \pm0.05  $ &$0.20  \pm0.06  $ & 1     \\
 B019D               & $9.09  \pm0.02  $ & $4.46  \pm0.10  $ &$0.30  \pm0.05  $ & 1     \\
 B035D               & $8.60  \pm0.03  $ & $4.26  \pm0.05  $ &$0.25  \pm0.09  $ & 1     \\
 B040-G102           & $7.76  \pm0.05  $ & $4.13  \pm0.10  $ &$0.25  \pm0.09  $ & 1     \\
\tableline
\end{tabular}
\end{center}
{$^a$ The reddening values are from \citet{kang12} ($\rm Flag=1, 2$)
and \citet{cald09, cald11} ($\rm Flag=3$).}
\end{table}

\end{document}